\documentclass{article}
\usepackage{graphicx}
\usepackage{latexsym,psfrag}
\usepackage{amssymb,amsfonts,amsmath}
\usepackage{epsf}
\usepackage[dvips]{color}
\title{
\begin{flushright}
{\small IFIC/06-16}
\end{flushright}
\vspace{1cm}
Critical exponents for higher-representation sources in 3D SU(3) gauge theory 
from CFT}
\author{Ferdinando Gliozzi$^a$ and Silvia Necco$^b$ }
\newcommand{\eq}{\begin{equation}}
\newcommand{\en}{\end{equation}}
\newcommand{\ear}{\begin{eqnarray}}
\newcommand{\rae}{\end{eqnarray}}
\newcommand{\Z}{\mathbb{Z}}

\newcommand{\uu}{\mathbb{I}}
\newcommand{\R}{{\cal R}}
\newcommand{\s}{{\cal S}}
\newcommand{\T}{{\cal T}}
\newcommand{\D}{{\cal D}}
\newcommand{\bra}{\langle}
\newcommand{\ket}{\rangle}

\newcommand{\tr}{{\rm tr}\,}
\newcommand{\Tr}{{\rm tr}^{~}}
\definecolor{M_Beige}         {rgb}{0.96 , 0.96 , 0.86}

\definecolor{M_Brown}         {rgb}{0.65 , 0.16 , 0.16}

\definecolor{M_Gold}          {rgb}{1.00 , 0.84 , 0.00}

\definecolor{M_LemonChiffon}  {rgb}{1.00 , 0.98 , 0.80}

\definecolor{M_Orange}        {rgb}{1.00 , 0.60 , 0.00}

\definecolor{M_Pink}          {rgb}{1.00 , 0.75 , 0.80}

\definecolor{M_Violet}        {rgb}{0.93 , 0.51 , 0.93}
\begin{document}
\maketitle
\noindent
$^a$Dipartimento di Fisica Teorica, Universit\`a di Torino and\\ INFN,
sezione di Torino, via P. Giuria, 1, I-10125 Torino, Italy.\\
\vskip .1cm
\noindent
$^b$IFIC - Instituto de F\'isica Corpuscular, 
Edificio Institutos de Investigaci\'on\\
Apartado de Correos 22085,  E-46071 Valencia - Spain\\
\vskip0.1cm
\begin{tabular}{rl}
e-mail:&gliozzi@to.infn.it  necco@ific.uv.es 
\end{tabular}
\begin{abstract}
We establish an exact mapping between the multiplication table of the 
irreducible representations of SU(3) and the fusion algebra of the 
two-dimensional  conformal 
field theory in the same universality class  of 3D SU(3) gauge theory at 
the deconfining point. In this way the  Svetitsky-Yaffe conjecture 
on the critical behaviour of Polyakov lines in the  fundamental representation 
 naturally extends to whatever representation one considers.
 As a consequence, the critical exponents of the 
correlators of these  Polyakov lines are determined. Monte Carlo simulations 
with sources in the symmetric two-index representation,  
combined with finite-size scaling analysis, compare very 
favourably  with these 
predictions.
\end{abstract}

\section{Introduction}
 To probe the structure of the vacuum of $SU(N)$ gauge theory we dispose 
of an infinite variety of external sources  transforming as arbitrary 
irreducible representations of the gauge group. One might wonder whether
the information extracted in this way is largely redundant, since 
the force acting on a  colour source  in a 
representation $\R$ built up of $j$ copies of the fundamental representation 
should depend only on its $N-$ality $k^{~}_\R\equiv j~({\rm mod} N)$, the 
reason being that all representations with same $k$ (hence transforming 
in the same manner under the center $\Z_N$) can be converted into 
each other by the emission of a proper number of soft gluons.  

Actually such a
property should be regarded as a feature of the IR limit, valid only when the 
source in question is very far from the other sources. In such a case this is 
subjected to a confining force only if the $N-$ality is non-vanishing. 
At intermediate scales 
lattice studies have shown long ago that this is not the case. Even  
sources in the adjoint representation (and therefore blind to the center)  
feel a linear rising potential, with a string tension larger than that 
of the fundamental representation \cite{aop}. Since then  most numerical 
experiments based on large Wilson loops 
\cite{cj,pt,sd,ba,cp} yield string tensions which depend on the specific 
representation $\R$ of the probe source rather than on its $N-$ality.
In the IR limit  the heavier 
$\R-$strings are expected to decay into the string with smallest string 
tension within the same $N$-ality class, called $k-$string.
A theoretical description of such a decay as a level-crossing phenomenon 
can be found in \cite{Gliozzi:2005dv}.
For a recent discussion on this subject see \cite{Shifman:2005eb}.
\par
A surprisingly similar problem emerges when considering the $SU(N)$ 
gauge model at the deconfining point. If the transition is second order, one 
obvious question concerns the critical behaviour of the Polyakov lines 
in arbitrary representations. Over the years, many studies have been 
dedicated to this subject  \cite{gw,dam,jk,ctd,dh,pi}. The 
well-verified Svetitsky- Yaffe (SY) conjecture \cite{sy} would place the 
finite-temperature $SU(N)$ gauge theory in the 
universality class of $\Z_N$ invariant spin model in one dimension 
less and with short-range interactions. There is a one-to-one correspondence 
between the irreducible representations of $\Z_N$ and  the $N-$ality 
values of $SU(N)$. Thus, one is tempted to conclude that  the non-abelian 
nature of $SU(N)$ and therefore whatever difference among sources in different 
representations with the same $N-$ality should be completely lost at 
criticality: if  only the global $\Z_N$ symmetry matters 
in characterising the universality class,       
there appears to be no room for independent exponents for Polyakov 
loops in different representations with the same $N-$ality. 
 
\par The surprising result is 
that sources in higher representations, according to various numerical 
experiments \cite{dam,ctd,dh}, correspond to different magnetisation 
exponents, one exponent for each representation. Actually a mean field 
approximation of the  effective $SU(2)$ Polyakov-line action at 
criticality in the $d\to \infty$ limit shows that the
leading amplitudes of higher representations vanish at strong coupling, 
and the sub-leading exponents become dominant, thus each higher representation 
source carries its own critical exponent\cite{dh}.  
\par
A somewhat similar conclusion has been reached in $SU(3)$ and $SU(4)$  
critical theories in $2+1$ dimensions, starting from a different point of view
\cite{Gliozzi:2005dv,Gliozzi:2005qe}: the SY conjecture leads to build up
 a map between the operator product expansion (OPE) of the Polyakov 
operators in the gauge theory and the corresponding spin operators in the 
two-dimensional conformal field theory (CFT) describing the associated 
spin system at criticality. An interesting property of such a correspondence
 is that one can define a conserved $N-$ality also on the CFT side and that 
the number of irreducible representations of the conformal algebra 
(and therefore the number of critical indices) is \emph{larger} than $N$, 
thus different Polyakov operators with the same $N-$ality can behave 
differently with an exactly prescribed critical index. 

In this paper 
we build up explicitly the above-mentioned  correspondence between the 
irreducible representations of SU(3) 
and the primary fields  of the CFT describing the critical behaviour of the 
(2+1) SU(3) gauge theory (see Sect.\ref{gaugeCFT}). This leads to some specific 
predictions on the critical indices to be associated to Polyakov loop operators 
in higher representations. 
 In Sect.\ref{MC} we compare  our predictions with a set of Monte Carlo 
simulations at criticality, using finite size scaling methods, finding a 
nice and complete agreement.

\section{The gauge/CFT correspondence}
\label{gaugeCFT}
The effective theory describing the interaction of Polyakov lines of any
$d+1$ gauge theory at a finite temperature $T$ can be described, 
when all the unrelated degrees of freedom are integrated out, as a 
$d-$dimensional spin model with a global symmetry coinciding with the 
center of the gauge group. Such an effective theory has only short-range 
interactions, as Svetisky and Yaffe (SY) observed long time ago \cite{sy}. 
As a consequence, if the deconfinement transition of the  gauge theory is 
second-order, it is in the same universality class of the order-disorder 
transition of the corresponding spin model.
\par
Therefore all the universal properties of the deconfining transition can be 
predicted  to coincide  with the ones  of the dimensionally reduced effective 
model. These include  the values of critical indices, the finite-size scaling 
behaviour, and the correlation functions at criticality.     
\par
When $d=2$ this SY conjecture, which passed several numerical tests, 
becomes particularly predictive because one can apply the  methods
of CFT.
\par
The main ingredient to fully exploit the predictive power of the SY conjecture 
is a mapping relating the physical observables of the gauge theory to the 
operators of the reduced model, as first advocated in  \cite{Gliozzi:1997yc}.
\par
The first entry of such a mapping is intrinsically contained in the SY 
conjecture, namely the correspondence between the order parameter of the 
deconfining transition of the gauge system, {\it i.e.} the Polyakov line 
in the fundamental representation $f$, and the order parameter $\sigma$ 
of spontaneous symmetry breaking transition  of the spin model:
\eq
\Tr_f(U_{\vec{x}}) ~\sim~ \sigma(\vec{x})~, 
\label{syf}
\en
$U_{\vec{x}}$
is the gauge group element  associated to the closed path 
winding once around the periodic imaginary time direction intersecting 
the spatial plane at the point $\vec{x}$. The above equivalence is only 
valid in a weak sense, that is, when the left-hand-side of the equation  
is inserted in a correlation function of the gauge theory and the right-hand 
side in the corresponding correlator of the spin model.
 \par One obvious question concerns the CFT operators corresponding
Polyakov lines in  higher representations. On the gauge side these are 
naturally generated by a proper combination of products of $\Tr_f U$, 
using repeatedly the  operator product expansion (OPE)
\eq
  \Tr_f(U_{\vec{x}})\; \tr_f(U_{\vec{y}})=
\sum_{\R\in f\otimes f}C_\R(\vert\vec{x}-\vec{y}\vert)\,\Tr_{\R}(U_{(\vec{x}+\vec{y})/2})
+\dots
\en
where $\R$ indicates any irreducible representation belonging in the 
decomposition of the direct product $f\otimes f$  and  the numerical 
coefficients 
$C_\R(r)$ are suitable functions; they become powers of $r$ at the critical 
point. The dots represent the contribution of higher dimensional 
local operators. For our purposes the relevant property of this kind of OPE is 
that the local operators  contributing to the right-hand-side  are classified 
according to the irreducible representations of the gauge group.
\par On the CFT side we have a similar structure.  The order parameter
$\sigma$ belongs to an irreducible representation $[\sigma]$ of the Virasoro 
algebra \cite{bpz} and the local operators contributing to an OPE are 
classified according to the decomposition of the direct product of the 
Virasoro representations of the left-hand-side operators. Such a decomposition 
is known as fusion algebra\cite{ve}  and can be written generically as
\eq
[\lambda_i]\star[\lambda_j]= c_{ij}^k[\lambda_k]
\label{fusion}
\en
where the non-negative integers $ c_{ij}^k$  are the fusion coefficients.
\par
The consistency of the SY conjecture requires a suitable mapping between the 
Clebsh-Gordan decomposition of the direct product of irreducible representations 
of the gauge group and the fusion algebra of the corresponding CFT.

\subsection{The $SU(3)$/3-state Potts-model Correspondence } 

In the present paper we are interested in the 2+1 dimensional $SU(3)$ 
gauge model which is described at the deconfinement point by the 
the same universality class of three-state Potts model, as it has been checked 
in numerical simulations \cite{ctd}.   
\par At the critical point such a model is described by a minimal CFT with central 
charge $c=\frac45$. The local operator content is composed by six primary 
fields $\lambda$ associated to six irreducible representations $[\lambda]$ that we list 
in Tab.\ref{tab:potts}  along with their scaling dimensions $x_\lambda$ and the Kac labels 
$(r,s)$ of the corresponding Virasoro representations.
\footnote{Actually the critical three-state Potts model is invariant under
a larger algebra than that of Virasoro, the so-called ${\cal W}_3$ algebra,
and the representations listed in Tab. \ref{tab:potts} are irreducible 
representations of such a larger algebra \cite{fz}. 
In fact the identity and the 
energy are the sum of two different irreducible Virasoro representations,  
as Tab. \ref{tab:potts} shows. For a recent discussion on the critical 
three-state Potts model see \cite{Caselle:2005sf}.}
\begin{table}[htb]
\caption{Operator content of CFT describing the critical three-state 
Potts model \label{tab:potts}}
\begin{center}  
\begin{tabular}{|c|cc|cc|cc|}
\hline
$\lambda$&$\sigma$ & $\psi$ & $\epsilon$ & $\uu$&$\sigma^+$&$\psi^+$\\
name&spin &simple &energy&identity&spin &simple \\
&field&current&&&field&current\\
\hline
& & & & & &\\[-0.2cm]
$x_\lambda$&$\frac2{15}$&$\frac43$&$\frac45$&$0$&$\frac2{15}$&$\frac43$\\[0.1cm]
$(r,s)$&$(3,3)$&$(4,3)$&$(2,1)\oplus(3,1)$&$(1,1)\oplus(4,1)$
&$(3,3)$&$(4,3)$\\[0.1cm]
3-ality&1&1&0&0&-1&-1\\[0.1cm]
$\D_\lambda$&$\frac{1+\sqrt{5}}2$&1&$\frac{1+\sqrt{5}}2$&1&$\frac{1+\sqrt{5}}2$
&1\\[0.1cm]

\hline
\end{tabular}
\end{center}
\end{table}

These are related to the scaling dimensions by
\eq
x_{r,s}=\frac{(6r-5s)^2-1}{60}~.
\en
The reason why we call $\psi$ a simple current will be clear in the next subsection.
The last two rows of Tab.\ref{tab:potts} refer to two important properties 
of the fusion algebra, which is presented in Tab.\ref{tab:fusion}. 

\begin{table}[htb]
\caption{Fusion algebra \label{tab:fusion}}
\begin{center}  
\begin{tabular}{|c|c|c|c|c|c|c|}
\hline
%\cline{2-7}
&$\sigma$ & $\psi$ & $\epsilon$ & $\uu$&$\sigma^+$&$\psi^+$\\[0.1cm]
\hline
& & & & & &\\[-0.2cm]
$\sigma$ &$\sigma^++\psi^+ $&$\sigma^+$&$\sigma+\psi$&
$\sigma$ &$\uu+\epsilon$
&$\epsilon$ \\[0.1cm]
$\psi$&$\sigma^+$&$\psi^+$&$\sigma$&$\psi$&$\epsilon$&$\uu$\\[0.1cm]
$\epsilon$&$\sigma+\psi$&$\sigma$&$\uu+\epsilon$&$\epsilon$&$\sigma^++
\psi^+$
&$\sigma^+$\\[0.1cm]
$\uu$&$\sigma$&$\psi$&$\epsilon$&$\uu$&$\sigma^+$&$\psi^+$\\[0.1cm]
$\sigma^+$&$\uu+\epsilon$&$\epsilon$&$\sigma^++\psi^+$&$\sigma^+$
&$\sigma+\psi$&$\sigma$\\[0.1cm]
$\psi^+$&$\epsilon$&$\uu$&$\sigma^+$&$\psi^+$
&$\sigma$&$\psi$\\[0.1cm]
\hline
\end{tabular}
\end{center}
\end{table}
First, these fusion rules show that the six representations may be cast into three 
doublets corresponding to the three  values ($1,0,-1$) of a  multiplicatively 
conserved quantum number, exactly like the triality in the $SU(3)$ gauge theory. 
This provides a non-trivial check of SY conjecture. 
\par Another useful notion is that 
of {\sl quantum dimension} $\D_\lambda$ associated to each representation $\lambda$  
of CFT and related in a 
profound way to the modular properties of the theory \cite{ve}. 
For our purposes  we simply define the $\D_\lambda$'s as the set of (real or complex) 
numbers obeying the sum rule
\eq
\D_\lambda\,\D_\mu=\sum_{\nu\in\lambda\star\mu}\D_\nu~,
\label{qdim}
\en 
which is a property that the quantum dimension shares with the  dimension 
$d_\R=\tr_\R{\bf 1}$ of a group representation $\R$: 
this is an integer satisfying the obvious relation
$d_\R\,d_\s=\sum_{\T\in\R\otimes\s}d_\T~$.
\par Applying such a definition to the fusion algebra of Tab.\ref{tab:fusion} 
one finds at once 
that the set of representations split into two parts; three of them have $\D=1$, 
while the quantum dimensions of the other 
three is a solution of the quadratic equation $x^2=x+1$, as listed in the last row of 
Tab.\ref{tab:potts}. The importance of such a notion  stems from the 
fact that the sum rule (\ref{qdim}) as well as  conservation of 
triality are sufficient conditions to fix uniquely the fusion algebra, as one can 
check at once.
\par We are now in a position to add other entries in the gauge/CFT mapping.   
It suffices to compare the fusion rules of Tab.\ref{tab:fusion} with the 
Clebsh-Gordan decomposition of the direct product of the corresponding $SU(3)$
irreducible representations or multiplets. These are labelled by a pair of 
 integers $(p,q)$ which give the numbers of covariant and contravariant indices.
The corresponding dimensions $d_{(p,q)}$ and  triality $k_{(p,q)}$ are given by 
\eq
d_{(p,q)}=(p+1)\,(q+1)\,(p+q+2)/2~,~~~~k_{(p,q)}\equiv\; (p-q)\; {\rm mod}~3~.
\label{dim}
\en   
To adhere to the standard notation we denote the SU(3) multiplets through their 
dimension $(p,q)\equiv \{d_{(p,q)}\}$. 
\par Comparison of the fusion rule
\eq
[\sigma]\star[\sigma^+]=[{\uu}]+[\epsilon]
\en
with the analogous one on the gauge side
\eq
\{3\}\otimes\{\bar{3}\}=\{1\}+\{8\}\,,~~{\rm or}~~f\,\otimes\,\bar{f}=1+adj
\label{b33}
\en
yields the new entry
\eq
\Tr_{adj}(U_{\vec{x}})~\sim~ a+\epsilon(\vec{x})~,
\label{cadj}
\en
that is expected to be valid for any $SU(N)$ gauge theory undergoing 
a continuous phase transition.
The constant $a$ can be numerically evaluated \cite{jk} using the 
expected finite-size behaviour  
\eq
\bra\Tr_{adj}(U)\ket\simeq a+\frac b{L^{2-1/\nu}}~, 
\en
where $L$ is the spatial size of the system. We used the general relation 
$x_\epsilon=d-1/\nu$ relating the scaling 
dimension of the energy operator to the thermal exponent  $\nu$. 
\par Similarly the fusion rule
\eq
[\sigma]\star[\sigma]=[\sigma^+]+[\psi^+]~,
\en
corresponds to 
\eq
\{3\}\otimes\{3\}=\{\bar{3}\}+\{6\}~,
\en
where the anti-fundamental representation $\{\bar{3}\}$ corresponds to the antisymmetric 
combination of two quarks while $\{6\}$ is the symmetric one.
Owing to the correspondence $\Tr_{\bar{f}}(U_{\vec{x}})\sim 
\sigma^+(\vec{x})$, the symmetric representation yields the new entry 
 \eq
\Tr_{\{6\}}(U_{\vec{x}}) ~\sim~ \psi^+(\vec{x})+
c\,\sigma^+(\vec{x})~.
\label{c6}
\en
The coefficient $c$ is necessarily different from zero: if it were 
vanishing, the fusion rule $[\psi^+]\star[\sigma]=[\epsilon]$ would correspond to
$\{6\}\otimes\{3\}=\{{8}\}+\{10\}$ which would in turn imply  either
 $a=0$ in Eq.(\ref{cadj}), which is not the case, or a dubious cancellation between
the contributions of rep.s $\{8\}$ and $\{10\}$.
As a consequence, the Polyakov-Polyakov critical correlator
of the symmetric representation $\{6\}$ is expected to have the 
following general form in the thermodynamic limit
\eq
\bra\Tr_{\{6\}}(U_{\vec{x}})\;\Tr_{\{\bar{6}\}}(U_{\vec{y}})
\ket=
\frac{c_s}{r^{2x_\sigma}}+\frac{c_u}{r^{2x_\psi}}~,
\label{c66}
\en
 with $r=\vert\vec{x}-\vec{y}\vert$ and $c_s,c_u$ suitable coefficients.
Since $x_\sigma<x_\psi$, the second term drops off more rapidly than the 
first, thus at large distance this correlator behaves like that of 
the anti-symmetric representation $\{\bar{3}\}$ as expected also at zero 
temperature.

\subsection{ A difficult question}
The SU(3)/CFT correspondence we have just established works in 
the sense that higher dimensional representations that can be screened 
to the fundamental are associated 
with operators that are suppressed in large distance correlators,
 like the $\{6\}$ decaying in $\{\bar3\}$ described by Eq. (\ref{c66}). A similar property 
emerges also  in SU(4) gauge theory \cite{Gliozzi:2005qe}. 

One might ask the following question:
how could the two-dimensional CFT "know" about the requirement of such a 
hierarchy of scaling dimensions due to the finite-T relation to 
non-abelian gauge theories? In other words, could one predict an ordering 
of scaling dimensions in 2D which agrees with the expectations of the 
corresponding gauge theory? \footnote{Actually this is a question posed 
to one of us by P. Damgaard.}. 
\par Since the core of the gauge/CFT 
correspondence  resides in the mapping between the Clebsh-Gordan 
decomposition of the direct product of representations of the gauge 
group and the CFT fusion algebra (\ref{fusion}), the  suggestion 
naturally arises  that the ordering of the scaling dimensions could be a 
direct consequence of the algebraic structure of the latter. 
\par Even if we failed 
in finding a complete proof of such a hierarchy of scaling dimensions, we still 
believe that this property should be somehow encoded in the fusion rules. 
Hints can be found in the relationship 
between fusion algebra and modular invariance we alluded in the Introduction,  
which yields restrictions on the allowed values of the scaling dimensions\cite{ve}. 
For instance, in a CFT 
with a finite number $M$ of primary fields, like in the case at hand, one gets
\cite{jf}
\eq
3\sum_{i=1}^{M}x_i=c\,\frac M4 {\rm mod}~\Z~,
\label{sumx}
\en 
where $c$ is the central charge. Further relevant information comes from 
the \emph{ simple currents}\cite{sch}. These are by definition those primary 
fields $\Psi$ which 
have unique fusion rules with all primaries of the theory, i.e.
\eq
\Psi\star\lambda_i=\lambda_j
\en 
with a single primary field $\lambda_j$ appearing on the right-hand side 
for any choice
of $\lambda_i$. Iterating the fusion rule of a simple current with itself 
generates a group $\Z_N$, with $\Psi^N=\uu$.  As an example, the primary 
$\psi$ of Tab.\ref{tab:fusion} is a simple current with $N=3$. The scaling 
dimensions of simple currents are strongly restricted. Monodromy properties
of the associated correlators easily yield 
\eq
x_{\Psi}^{~}=m\,\frac{N-1}{N}\,{\rm mod}~\Z~,
\en
with some integer $m$ modulo $N$. This condition restricts the scaling 
dimensions of $\psi$ to two possible values $x_\psi=\frac23$ or 
$\psi=\frac43$ (which is the correct value). Inserting this constraint into 
Eq(\ref{sumx}) we get the further restriction $x_\epsilon+2\,x_\sigma=\frac15+\frac n3$, 
with $n$ arbitrary integer, which does not suffices to find the complete solution.

\section{Monte Carlo simulations}
\label{MC}
 We performed Monte Carlo simulations on the finite-T (2+1)-dimensional 
SU(3) gauge model with standard plaquette action at the deconfinement point, using two 
critical couplings estimated in \cite{Engels:1996dz}, namely $\beta_c=8.155(15)$  for the 
temporal extension $N_t=2$ and $\beta_c=14.74(5)$ at $N_t=4$ while the spatial 
size $L$ of the lattice was chosen in the range $8\le L\le 64$. \\
We adopted an hybrid updating algorithm consistent of one heat-bath step
combined with $N_{or}$ over-relaxation steps; in particular we used
$N_{or}=10$ in all our simulations \footnote{Notice that we did not
  perform a systematic study in order to optimise the choice of $N_{or}$.}.\\
It is well known that in the confined phase expectation values of large Wilson loops and Polyakov loops correlation functions at large distances are difficult to measure, since the signal-to-noise ratio decreases exponentially.
For this reason, efficient variance reductions methods have been 
developed \cite{lw}. At the deconfining point the situation is much more 
favourable because the exponential decay is replaced by a power law; 
correlation functions at large distances can be measured with good precision 
without employing special techniques.\\
For the different extensions $L$ at $N_t=2$ we collected between 10000 and
30000 measurements. For $N_t=4$ we collected 10000 measurements for each $L$ 
Between two measurements we performed 100-200 updating steps. 
At the critical point update algorithms suffer from critical slowing down and 
one expects large autocorrelation times of the observables; we analysed our
data through jackknife binning by using sufficiently large sizes of the bins in order
to take this into account.

\subsection{Finite-size scaling analysis}
 Observing the second term of Eq.(\ref{c66}) is very challenging
from a computational point of view, because it drops off much more rapidly than 
the fist term, being $x_\psi$  rather large 
(actually $x_\psi=10\,x_\sigma$). In order to gain a better control of this behaviour
one has to resort to finite size scaling analysis.

The critical two-point function of the spin field $\sigma$ in a torus $L,L'$, i.e. in 
a rectangle  with periodic boundary conditions of periods $L$ and $L'$ in the $x$ and $y$ 
directions respectively, has the finite-size scaling form
\eq
\bra\sigma(0,0)\sigma^+(x,y)\ket=L^{-2x_\sigma}f(x/L,y/L')~,   
\label{torus}
\en
where $f$ is a universal function. Unfortunately the general Ansatz for 
CFT correlation functions on a torus proposed long ago \cite{bnz} cannot be applied to 
the present case, therefore the form of $f$ for the 
3-state Potts model is substantially unknown, apart from the limit
$L'\to\infty$ (cylindrical geometry), where the principle of conformal invariance at the 
critical point allows to write the explicit, exact form of $f$ in any CFT \cite{cardy}.
\par In our numerical simulations we put instead $L=L'$.
 Note that for $r=\sqrt{x^2+y^2}\ll L$ $f$ becomes a function of the single variable 
$\frac rL$. In the thermodynamic limit $L\to\infty$ scaling considerations yield
\eq
f(x/L,y/L)\to\, \left(\frac Lr \right)^{2x_\sigma}~.
\en
\par
\begin{figure}
\centering
{
\psfrag{r/L}{$r/L$}
\psfrag{cor}{$\bra {\rm tr}_fU_0\,{\rm tr}_fU^\dagger_{\frac rL}\ket
(\frac{L}{L_o})^{2x_\sigma}$}
\psfrag{20 +}{\textcolor{red}{$L=20~~+$}}
\psfrag{24 x}{\textcolor{green}{$L=24~~\times$}}
\psfrag{32 T}{\textcolor{blue}{$L=32~~\star$}}
\psfrag{64 o}{\textcolor{magenta}{$L=64~~\bar{\sqcup}$}}
{\includegraphics[width=12.cm]{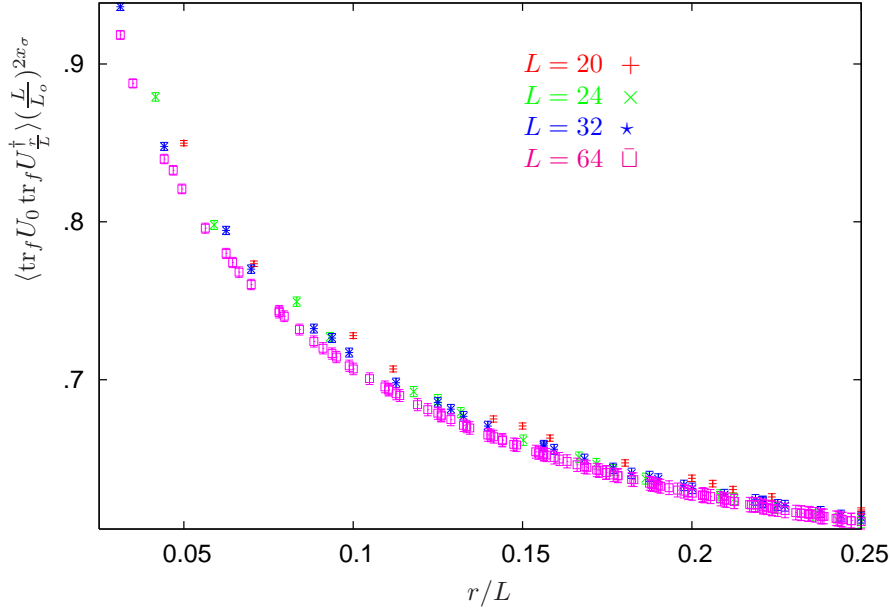}}
}
\caption{Test of the finite-size scaling form (\ref{foruss})  on a $L\times L\times 2$ 
lattice. $L_o$ is a reference scale.
\label{Figure:1}}
\end{figure}

 The SY conjecture predicts that, combining (\ref{syf})and (\ref{torus}),
the Polyakov-Polyakov  correlator in fundamental representation, defined as 
\eq
{\cal G}_{\{3\}\,}(x,y)\equiv \bra \tr_{\{3\}}U_{(0,0)}\;
\tr_{\{{3}\}}U^\dagger_{(x,y)}\ket
\en
should behave as
\eq
{\cal G}_{\{3\}}(x,y) \simeq
L^{-2x_\sigma}f(\xi_1,\xi_2)~,~~\xi_1=x/L~,\,\xi_2=y/L~.   
\label{forus}
\en
It should be stressed that, in writing Eq.(\ref{forus}), as well as 
the analogous ones (\ref{mix}) and (\ref{L66}) below, as with all lattice 
identifications of scaling operators, the correlators of either side are asymptotically
equal  only when $L$ is large and  the points are far apart. At smaller 
separations  there are additional, less relevant operators on the right-hand 
side which 
will give rise to corrections. These are visible in Fig. \ref{Figure:1} where we plotted
the quantity $\bra \tr_{\{3\}} U_0\,\tr_{\{3\}}U^\dagger_{\frac rL}\ket
(\frac{L}{L_o})^{2x_\sigma}$ versus $r/L$. The points fall with good accuracy on a single 
curve - thus verifying the Ansatz (\ref{forus})- only for $r/L> 0.15$.
The first correction to scaling of (\ref{forus}) is a term proportional to
$L^{-2x_\sigma-2}$: 
\eq
{\cal G}_{\{3\}}(x,y)
\simeq
L^{-2x_\sigma}f(\xi_1,\xi_2)+ L^{-2x_\sigma-2}\bar{f}(\xi_1,\xi_2)~,  
\label{foruss}
\en
where $\bar f$ is another scaling function.
Other corrections come from the fact that the lattice system is not 
exactly at the critical point, however  there is no sign of this kind of 
corrections within the accuracy of our data.  

 \begin{figure}[thb]
\centering
{
\psfrag{L}{$L$}
\psfrag{correlators}{{\sl correlators}}
{\includegraphics[width=12.cm]{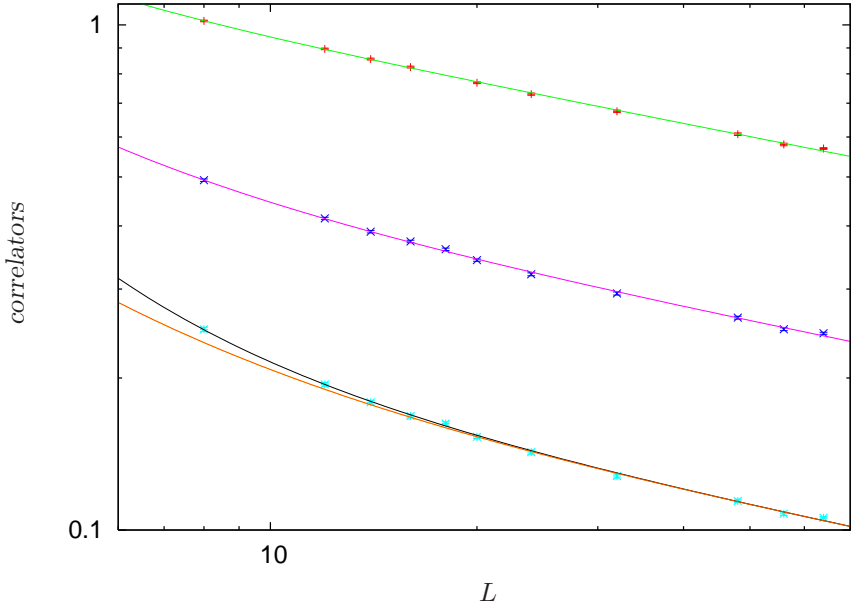}}
}
\caption{ Finite-size scaling analysis of the Polyakov-Polyakov
correlators at a distance $r=L/2$ in a $L\times L\times2$ lattice as a 
function of the size $L$.
 The top line is a fit of $\bra \tr_{\{3\}}U_{(0,0)}\;\tr_{\{{3}\}}
U^\dagger_{(0,\frac L2)}\ket$ data  
to (\ref{foruss}); it is apparently a straight-line in this log log plot,
indicating that the corrections to scaling are small; the medium curve is a 
fit  of 
$\bra \tr_{\{3\}}U_{(0,0)}\;\tr_{\{6\}}U_{(0,\frac L2)}\ket$  to (\ref{mix}). 
 The bottom line is not a fit: it is determined by
 the other two sets of data by putting $h=0$ in (\ref{L66}). The short distance data 
show the non-vanishing contribution of the $\bra\psi\,\psi^+\ket$ correlator. 
\label{Figure:2} }
\end{figure}

\begin{table}[thb]
\caption{Estimates of the universal functions $f(\xi_1,\xi_2)$ and  $g(\xi_1,\xi_2)$ 
and of the constant $c$ in critical SU(3) lattice models with temporal size 
$N_t=2$\label{tab:Nt2}}
\begin{center}
\begin{tabular}{|cc|c|c|c|}
\hline
& & & &\\[-0.2cm]
$i$&$j$&$f(i/4,j/4)/f(1/4,1/4)$&$g(i/4,j/4)/g(1/4,1/4)$&$c$\\[0.1cm]
& & & &\\[-0.2cm]
\hline
& & & &\\[-0.2cm]
1&0& 1.8403(41)/$f(1/4,1/4)$& 1.556(86)/$g(1/4,1/4)$&0.4273(15)\\[0.1cm]
1&1&0.9616(44)&0.91(10)&0.4257(15)\\[0.1cm]
2&0&0.9485(44)&0.89(10)&0.4255(17)\\[0.1cm]
2&1&0.9369(45)&0.83(10)&0.4255(17)\\[0.1cm]
2&2&0.9268(44)&0.79(10)&0.4260(18)\\[0.1cm]
\hline
\end{tabular}
\end{center}

\end{table}

\begin{table}[thb]
\caption{Estimates of the universal functions $f(\xi_1,\xi_2)$ and  $g(\xi_1,\xi_2)$ 
and of the constant $c$ in critical SU(3) lattice models with temporal size 
$N_t=4$\label{tab:Nt4}}
\begin{center}
\begin{tabular}{|cc|c|c|c|}
\hline
& & & &\\[-0.2cm]
$i$&$j$&$f(i/4,j/4)/f(1/4,1/4)$&$g(i/4,j/4)/g(1/4,1/4)$&$c$\\
& & & &\\[-0.2cm]
\hline
& & & &\\[-0.2cm]
1&0& 1.5596(70)/$f(1/4,1/4)$& 2.50(30)/$g(1/4,1/4)$&0.3113(30)\\[0.1cm]
1&1&0.9597(85)&0.87(19)&0.3103(25)\\[0.1cm]
2&0&0.9468(81)&0.84(19)&0.3097(24)\\[0.1cm]
2&1&0.9360(77)&0.79(17)&0.3102(23)\\[0.1cm]
2&2&0.9264(78)&0.77(18)&0.3096(25)\\[0.1cm]
\hline
\end{tabular}
\end{center}

\end{table}

\par As a next step we considered the  mixed correlator 
$\bra\sigma(0,0)\,\psi^+(x,y)\ket$. 
This is zero in the thermodynamic limit 
because the fusion rule 
$[\sigma]\star[\psi^+]=[\epsilon]$ does not contain the identity $[\uu]$. On a torus 
it can be expressed in terms of another universal function
\eq
 \bra\sigma(0,0)\psi^+(x,y)\ket=L^{-x_\sigma-x_\psi}g(\xi_1,\xi_2)~.
\label{sipsi}
\en
Again the scaling dimensions of the involved operators  imply, for large $L$,  
\eq
 g(x/L,y/L)\to \left(\frac Lr\right)^{x_\sigma+x_\psi-x_\epsilon}~.
\en
 Therefore the new entry (\ref{c6}) of the SY conjecture yields  
\eq
\bra \tr_{\{3\}}U_{(0,0)}\,\tr_{\{{6}\}}U_{(x,y)}\ket\simeq
c\, {\cal G}_{\{3\}}(x,y)  +L^{-x_\sigma-x_\psi}g(\xi_1,\xi_2)~.
\label{mix}
\en
The finite-size scaling form of the  simple current correlator 
$\bra\psi(0)\,\psi^+(\vec{x})\ket$ is associated to a third universal function
$h(x/L,y/L)$
\eq
\bra\psi(0,0)\psi^+(x,y)\ket=L^{-2x_\psi}h(\xi_1,\xi_2)~,   
\label{porus}
\en
with 
\eq
h(x/L,y/L)\to\, \left(\frac Lr\right)^{2x_\psi}
\en
in the large volume limit.
Hence, on the gauge side, the Polyakov correlator in the $\{6\}$ 
representation can be written as 
\eq
\bra \tr_{\{6\}}U_{(0,0)}\tr_{\{{6}\}}U^\dagger_{(x,y)}\ket\simeq
c^2\,{\cal G}_{\{3\}}(x,y) +
2c\frac{g(\xi_1,\xi_2)}{L^{x_\sigma+x_\psi}}+\frac{h(\xi_1, \xi_2)}{L^{2x_\psi}}~.
\label{L66}
\en
\par In order to test these finite-size scaling relations it is worth noting that a 
rescaling of both the lattice size and the distance $r$ by a common factor $s$ is 
compensated, at criticality, by a rescaling of the correlation function which depends 
on the scaling dimension of the involved operators, as Eq.s (\ref{forus}), (\ref{mix}) and 
(\ref{L66}) clearly show.

 In practice we proceeded as follows. We chose both $x$ and $y$  
of the form $ jL/{4}$ with $j=0,1,2$.\footnote{In this way the minimal 
non-vanishing distance considered $(r=L/4)$ lies in the scaling region inferred from 
Fig.\ref{Figure:1}.} Varying  the linear size $L$ over a set of different values
\footnote{Actually the number of different lattice sizes was 10 for $N_t=2$ and 6 for 
$N_t=4.$}, we generated, for each choice 
of $r/L$, a sample of data. Since different lattice sizes imply different numerical 
experiments, our data are by construction statistically independent. A typical set of 
data is shown in Fig.\ref{Figure:2}.
In all the cases we do not 
attempted to use the scaling dimensions as fitting parameters. The $\chi^2$ values of 
such a kind of fits provide a check on the systematic errors arising from corrections 
to scaling. In all the fits the $\chi^2/d.o.f$ was in the range between  1 and 2.

Estimates of the universal functions $f(\xi_1,\xi_2) $ and $g(\xi_1,\xi_2)$ in the five 
special points considered, as well as the constant $c$ are reported in Tab.\ref{tab:Nt2}
$(N_t=2)$ and in Tab.\ref{tab:Nt4}  $(N_t=4)$. We consider the fact that these functions 
on lattices with different temporal extensions are substantially the same within 
the errors, apart from a different multiplicative normalisation, as a highly non-trivial 
consistency test of the present description. It is also remarkable the $c$ evaluated 
on different points $(\xi_1,\xi_2)$ has a constant and stable value, as  (\ref{mix}) 
and (\ref{L66}) require. 
\subsection{Multiplets of vanishing triality}

In a SU(N) gauge theory at finite temperature one could choose as the exact  order 
parameter of the deconfinement transition the Polyakov line in any representation of 
non-zero $N$-ality. At the deconfining point we have
\eq
\lim_{L\to\infty}\,\bra\tr_\R U\ket_{T=T_c}=0 ~,\forall\, k_\R\not=0~. 
\en
One obvious question concerns the critical behaviour of Polyakov lines corresponding to 
sources in  representations
of zero $N$-ality.

A straightforward consequence our gauge/CFT correspondence is that the  finite- 
size behaviour (\ref{cadj}) of the adjoint representation can be enlarged to any multiplet 
of vanishing $N$-ality. In the present SU(3) case we can use our Monte Carlo 
data at $r=0$ to extract the vacuum expectation value for the 0-triality multiplets 
of dimensions 8,10 and 27. Indeed, combining (\ref{b33}) with the 
Clebsh-Gordan decompositions
\eq
\{3\}\otimes\{6\}=\{8\}+\{10\}~;\;
\{\bar6\}\otimes\{6\}=\{1\}+\{8\}+\{27\}~,
\en  
and using the new entry (\ref{c6}), we get
\eq
\bra\Tr_{\R_o}U\ket=a_{\R_o}^{~}+b_{\R_o}^{~}\,{L^{-x_\epsilon}}+\ldots~
\R_o=\{8\}\,,\{10\}\,,\{27\}~,
\label{zeroN}
\en
where the ellipses indicate the corrections to scaling \footnote{The first two corrections 
are the first descendent field of $[\uu]$ which gives a contribution 
$\propto L^{-2}$ and the subleading thermal operator $\epsilon'$, corresponding to the 
Kac labels $(3,1)$ of Tab. \ref{tab:potts}, which gives a contribution
$\propto L^{-x_{\epsilon'}}$ with $x_{\epsilon'}=\frac{14}5$.}. In our case they can be 
neglected, being the $\chi^2/d.o.f$ values of the fits to (\ref{zeroN}) always less
than 1. The estimates of the parameters $a_{\R_o}^{~}$ and $b_{\R_o}^{~}$
are reported in Tab.\ref{tab:zeroN}. 
\begin{table}[thb]
\caption{Fits to  Eq.(\ref{zeroN}) for various representations of zero triality
\label{tab:zeroN}}
\begin{center}
\begin{tabular}{|c|ccc|ccc|}
\hline
& & & & & &\\[-0.3cm]
$\R_o^{~}$&$N_t$&=&2&$N_t$&=&4\\
&$a_{\R_o}^{~}$&&$b_{\R_o}^{~}$&  $a_{\R_o}^{~}$&&$b_{\R_o}^{~}$\\[0.1cm]
& & & & & &\\[-0.3cm]
\hline
& & & & & &\\[-0.3cm]
$\{8\}$ &0.5640(10) &&0.985(10)&0.3589(13)&&1.218(26)\\[0.1cm]
$\{10\}$&0.0804(20)&&0.221(25)&0.3144(26)&&0.160(52)\\[0.1cm]
$\{27\}$&0.0592(21)&&0.181(27)&0.0167(24)&&0.101(49)\\[0.1cm]
\hline
\end{tabular}
\end{center}

\end{table}
Note that the quantities $a_{\R_o}^{~}$ represent the thermodynamic limit of the vacuum
 expectation value of the Polyakov line in the representation $\R_o$. It turns out that 
representations of zero $N$-ality yield $a_{\R_o}^{~}\not=0$.  Similar results have 
also been  found in 4D gauge theories near the deconfining point \cite{dh,pi}.
  
\section{Conclusion}
In this paper we presented  a consistent extension 
of the Svetitsky-Yaffe conjecture in (2+1) dimensional
SU(3) gauge theory at finite temperature which led us  to associate  
Polyakov lines in arbitrary
representations of  the gauge group to suitable conformal operators of the
corresponding 2D CFT. 

In particular, we  built up a correspondence between the multiplication table of the 
irreducible representations of the gauge group and the fusion algebra of the 
primary operators of the critical 3-state Potts model. 

One important consequence is that the critical exponents of the 
correlators of these  Polyakov loops are univocally determined. We  studied in 
particular the 
Polyakov line in the symmetric, two-index representation $\{6\}$ and determined the 
functional finite-size form of some related correlator. We also discussed 
the vacuum expectation value of a single Polyakov line in the first few multiplets
of vanishing triality. We tested these predictions in 
high precision Monte Carlo simulations finding complete agreement. 
\vskip .5 cm

{\bf Acknowledgements:} 
FG would like to thank P. Damgaard for fruitful correspondence. SN would like
to thank the PH-TH division at CERN for the hospitality during the completion
of this work. We thank moreover CPT Marseille and IFIC Valencia for computing resources.


\begin{thebibliography}{9}
\bibitem{aop}  C. Bernard, Nucl Phys. B 
{\bf 219} (1983) 341.

J. Ambj\o rn, P.Olesen and C. Peterson, Nucl Phys. B 
{\bf 240},186 (1984; Nucl. Phys. B {\bf 240}, 533 (1984).
\bibitem{cj} N.A. Campbell, I.H. Jorysz, C. Michael, Phys. Lett.B {\bf 167}
(1986) 91. 
\bibitem{pt} G. Poulis and H.D. Trottier, Phys. Lett. B {\bf 400}(1997)358
[arXiv:hep-lat/9504015];  C.  Michael, ArXiv:hep-lat/9809211.
\bibitem{sd} S. Deldar, Phys. Rev. D {\bf 62}(2000) 034509 
[arXiv:hep-lat/9911008]
\bibitem{ba}G. Bali, Phys. Rev. D {\bf 62} (2000) 114503 
[arXiv:hep-lat/0006022]. 
\bibitem{cp} C. Piccioni, arXiv:hep-lat/0503021.
\bibitem{Gliozzi:2005dv}
  F.~Gliozzi, 
%The decay of unstable k-strings in SU(N) gauge theories 
%at zero and finite temperature'
JHEP {\bf 08} (2005)063 
[arXiv:hep-th/0507016].

\bibitem{Shifman:2005eb}
  M.~Shifman,
  %``k strings from various perspectives: QCD, lattices, string theory and toy
  %models,''
  Acta Phys.\ Polon.\ B {\bf 36} (2005) 3805
  [arXiv:hep-ph/0510098].

\bibitem{gw} M. Gross and J. F. Wheater, Phys. Rev. Lett. {\bf 54} (1985) 389;
P. H. Damgaard , Phys. Lett. B {\bf 183}(1987) 81;
M.E.Faber, H. Markum and M. Meinahart, Phys. Rev. D {\bf 36} (1987)
632.
\bibitem{dam} P.H. Damgaard, Phys. Lett. B {\bf 194} (1987) 107;
K. Redlich and H. Satz, Phys. Lett. B {\bf 213} (1988) 191;
J. Christensen and P.H.Damagaard, Nucl. Phys. B {\bf 348}
(1991) 226; Nucl. Phys. B {\bf 354} (1991) 339.
\bibitem{jk} J. Kiskis, Phys. Rev. D {\bf 41} (1990) 3204; J.Kiskis and 
P. Vranas, Phys. Rev. D {\bf 49} (1994) 528.
\bibitem{ctd} Christensen, G. Thorleifsson, P.H.Damagaard and 
J.F. Wheater, Nucl. Phys. B{\bf 374} (1992) 225.
\bibitem{dh} P.H. Damgaard and M. Hasenbusch, Phys. Lett. B 
{\bf 331} (1994) 400.

\bibitem{pi}
  A.~Dumitru, Y.~Hatta, J.~Lenaghan, K.~Orginos and R.~D.~Pisarski,
  %``Deconfining phase transition as a matrix model of renormalized Polyakov
  %loops,''
  Phys.\ Rev.\ D {\bf 70} (2004) 034511
  [arXiv:hep-th/0311223].
\bibitem{sy}B. Svetitsky and L.G. Yaffe, Nucl. Phys. B {\bf 210} [FS6]
(1982) 423.
\bibitem{Gliozzi:1997yc}
  F.~Gliozzi and P.~Provero,
  %``The Svetitsky-Yaffe conjecture for the plaquette operator,''
  Phys.\ Rev.\ D {\bf 56} (1997) 1131
  [arXiv:hep-lat/9701014].
  %%CITATION = HEP-LAT 9701014;%%

\bibitem{Gliozzi:2005qe}
  F.~Gliozzi,
  %``Screening of sources in higher representations of SU(N) gauge theories at
  %zero and finite temperature,''
  PoS {\bf LAT2005} (2005) 196
  [arXiv:hep-lat/0509085].
  %%CITATION = HEP-LAT 0509085;%%


\bibitem{bpz} A.A. Belavin, A.M. Polyakov and  A. B. Zamolodchikov, 
Nucl. Phys. B {\bf 241} (1984) 333.
\bibitem{ve} E.Verlinde, Nucl. Phys. B {\bf 300} (1988) 360.
\bibitem{fz}
V.A. Fateev and A.B. Zamolodchikov, Nucl. Phys. B {\bf 280}
(1987) 644.
\bibitem{Caselle:2005sf}
  M.~Caselle, G.~Delfino, P.~Grinza, O.~Jahn and N.~Magnoli,
  %``Potts correlators and the static three-quark potential,''
  J.\ Stat.\ Mech.\  {\bf 0603} (2006) P008
  [arXiv:hep-th/0511168].
\bibitem{jf}J. Fuchs, Affine Lie Algebras and Quantum Groups, Cambridge 
University press, 1992.
\bibitem{sch}A.N. Schellekens and S. Yankielowicz,
 Nucl Phys. B{\bf 327} (1989) 673.
%\cite{Engels:1996dz}
\bibitem{Engels:1996dz}
  J.~Engels, F.~Karsch, E.~Laermann, C.~Legeland, M.~Lutgemeier, 
B.~Petersson and T.~Scheideler,
  %``A study of finite temperature gauge theory in (2+1) dimensions,''
  Nucl.\ Phys.\ Proc.\ Suppl.\  {\bf 53} (1997) 420
  [arXiv:hep-lat/9608099].
\bibitem{lw} M. L\"uscher and P. Weisz, JHEP {\bf 0109} (2001) 013 
\bibitem{bnz}J.~Bagger, D.~Nemeschansky and J.~B.~Zuber,
  %``Minimal Model Correlation Functions On The Torus,''
  Phys.\ Lett.\ B {\bf 216} (1989) 320.
\bibitem{cardy} J. Cardy, J.Phys. A {\bf 17} (1984) L385.
\end{thebibliography}
\end{document}